# CONTACT PAIRING INTERACTION FOR THE HARTREE-FOCK-BOGOLIUBOV CALCULATIONS

J. DOBACZEWSKI

*Institute of Theoretical Physics, Warsaw University, Hoża 69, PL-00-681 Warsaw, Poland*

W. NAZAREWICZ

*Institute of Theoretical Physics, Warsaw University, Hoża 69, PL-00-681 Warsaw, Poland*

*Department of Physics & Astronomy, University of Tennessee, Knoxville, Tennessee 37996, USA*

*Physics Division, Oak Ridge National Laboratory, Oak Ridge, Tennessee 37831, USA*

AND

M.V. STOITSOV

*Institute of Nuclear Research and Nuclear Energy, Bulgarian Academy of Sciences, Sofia-1784, Bulgaria*

**Abstract.** Properties of density-dependent contact pairing interactions in nuclei are discussed. It is shown that the pairing interaction that is intermediate between surface and volume pairing forces gives the pairing gaps that are compatible with the experimental odd-even mass staggering. Results of the deformed Hartree-Fock-Bogoliubov calculations for this "mixed" pairing interaction, and using the transformed harmonic oscillator basis, are presented.

## 1. Introduction

Recent advances in radioactive ion beam technology have opened up the possibility of exploring very weakly bound nuclei in the neighborhood of the particle drip lines [1, 2, 3, 4]. A proper theoretical description of such weakly bound systems requires taking into account the particle-particle (p-p, pairing) correlations on the same footing as the particle-hole (p-h) correlations, which - on the mean-field level - is done in the framework of the Hartree-Fock-Bogoliubov (HFB) [5, 6] or relativistic Hartree-Bogoliubov (RHB) [7] theories. In these methods, it is essential to solve the equations for the self-consistent densities and mean fields in the coordinate representation in order to allow the pairing correlations to build up with a full coupling to particle continuum. This task can be easily accomplished when the spherical symmetry is imposed; however, for deformed systems the problem becomes very difficult. The coordinate-space code working in the limit of the axial symmetry became available only very recently [8], while the code which is applicable also to triaxial deformations [9] is able to take into account only a very limited part of the positive-energy phase space [10, 11].

As an alternative method, the partial differential HFB equations have recently been solved for both spherical and axially deformed nuclei by expanding the quasiparticle wave functions in a complete set of so-called transformed harmonic oscillator (THO) single-particle wave functions [12, 13, 14]. They are derived from the standard harmonic oscillator basis by the unitary local-scaling coordinate transformation [15, 16, 17] which preserves many useful properties of the harmonic oscillator wave functions, and, in addition, it gives us access to the precise form dictated by the desired asymptotic behavior of the HFB densities. The resulting configurational HFB+THO calculations present a promising alternative to algorithms that are being developed for a coordinate-space solution to the HFB equations.

Apart from developing proper theoretical tools to solve the HFB or RHB equations, one has to choose appropriate nuclear effective forces responsible for the description of the weakly bound systems. Concerning the p-h channel, a variety of effective forces such as, e.g., the Skyrme and Gogny interactions [18], or interactions based on relativistic Lagrangians [7], have been extensively applied in the study of drip-line systems. In the p-p channel, the finite-range Gogny interaction and the zero-range delta interactions have been used. The pairing forces used in the p-p channel have been adjusted to properties of nuclei close to the stability line. Unfortunately, for drip-line nuclei, in which the pairing effects are crucially important due to the coupling to the continuum, the effective pairing interaction is not known.

Since in finite nuclei no derivation of the pairing force from first principles is available yet, there are many variations in the choice of pairing forces used in HFB and RHB calculations. When using the Gogny effective interaction in the p-h channel, the most "natural" choice is to parameterize the pairing force by the same finite-range Gogny force [19, 20, 21, 22]. Obviously, the same "natural" choice for the pairing force is a contact delta interaction when used in combination with the effective Skyrme forces [5, 6]. (The "natural" choice does not have to be the right one. Microscopically, the effective p-p interaction does not have to be the same as the in-medium p-h force [23].) In the case of relativistic approaches, both types of pairing are used in combination with effective delta-like p-h interactions based on relativistic Lagrangians [7, 24].

In the present paper we analyze the coordinate-space spherical HFB results for semi-magic nuclei and discuss properties of the zero-range pairing interaction. New "intermediate-type" pairing is suggested that takes an intermediate position between volume and surface delta pairing usually applied. With such a pairing force, we carry out the full HFB+THO mass-table calculations for even-even axially deformed nuclei.

## 2. Hartree-Fock-Bogoliubov theory

Within the HFB theory [18] the wave function of the even-even system is approximated by a generalized product state that represents the quasi-particle vacuum. This wave function is defined in terms of the amplitudes $(U, V)$ obtained by solving the HFB equations:

$$
\begin{pmatrix} h - \lambda & \Delta \\ -\Delta^* & -h^* + \lambda \end{pmatrix} \begin{pmatrix} U_n \\ V_n \end{pmatrix} = E_n \begin{pmatrix} U_n \\ V_n \end{pmatrix} , \tag{1}
$$

where $E_n$ are the quasiparticle energies, $\lambda$ is the chemical potential, and the matrices $h \, (= t + \Gamma)$ and $\Delta$ are defined by the matrix elements of the two-body interaction:

$$
\begin{aligned}
\Gamma_{\alpha\alpha'} &= \sum_{\beta\beta'} \bar{v}_{\alpha\beta\alpha'\beta'} \rho_{\beta'\beta}, \\
\Delta_{\alpha\alpha'} &= \tfrac{1}{2} \sum_{\beta\beta'} \bar{v}_{\alpha\alpha'\beta\beta'} \kappa_{\beta\beta'}.
\end{aligned} \tag{2}
$$

The chemical potential $\lambda$ has to be determined (separately for neutrons and for protons) by the subsidiary particle-number conditions.

## 3. Truncation of the finite configuration space

Expressions for the density matrix $\rho$ and the pairing tensor $\kappa$ [18] in terms of the HFB amplitudes $U(E_n, \mathbf{r})$ and $V(E_n, \mathbf{r})$,

$$
\begin{aligned}
\rho(\mathbf{r}, \mathbf{r}') &= \sum_{0 \leq E_n \leq E_{\max}} V^*(E_n, \mathbf{r}) V(E_n, \mathbf{r}') , \\
\kappa(\mathbf{r}, \mathbf{r}') &= \sum_{0 \leq E_n \leq E_{\max}} V^*(E_n, \mathbf{r}) U(E_n, \mathbf{r}') ,
\end{aligned} \tag{3}
$$

invariably require a truncation of the quasiparticle eigenstates by defining a cut-off quasiparticle energy $E_{\max}$, and then including all quasiparticle states only up to this value. When the finite-range Gogny force is used in the p-p channel, the cut-off energy $E_{\max}$ has only numerical significance. In contrast, the HFB calculations that use the zero-range pairing force require a finite space of states; otherwise, they give divergent energies with increasing $E_{\max}$ (see discussion in Ref. [6]). Recently, a regularization scheme has been devised that allows putting the cut-off prescription on firm theoretical ground [25, 26].

In practice, an efficient cut-off procedure can be devised [14] by using the so-called equivalent HFB single-particle spectrum $\bar{e}_n$ [5], defined by:

$$\bar{e}_n = (1 - 2N_n)E_n, \tag{4}$$

where $N_n$ denotes the norm of the lower HFB wave function. This spectrum is usually also used in the HFB calculations to readjust the values of the neutron (proton) chemical potential to obtain the correct values of the neutron (proton) particle number.

Due to a similarity between the equivalent energies $\bar{e}_n$ and the single-particle energies $e_n$, one may use the former ones to define an appropriate cut-off procedure for the quasiparticle states. Note that each equivalent energy characterizes a single given quasiparticle state, while the single-particle energies do not obey such a direct relationship. Therefore, the cut-off is realized by taking into account only those quasiparticle states for which $\bar{e}_n \leq \bar{e}_{\max}$, where $\bar{e}_{\max} > 0$ is a parameter defining the amount of the positive-energy phase space taken into account. All the hole-like quasiparticle states correspond to $N_n < 1/2$, and hence they have negative values of $\bar{e}_n$. Therefore, the condition $\bar{e}_n \leq \bar{e}_{\max}$ guarantees that they are all taken into account, even if they correspond to high (positive) quasiparticle energies. In this way, a global cut-off prescription is defined which fulfills the requirement of taking into account the positive-energy phase space as well as all quasiparticle states up to the highest hole-like quasiparticle energy.

## 4. Density-dependent contact pairing forces

In the actual HFB calculations based on the Skyrme forces in the p-h channel (as, e.g., the SLY4 parametrization [27] used in the present work), contact pairing interaction is usually used in the p-p channel. Two different forms have been used up to now – the volume type,

$$V^{\delta}_{\mathrm{vol}}(\mathbf{r}, \mathbf{r}') = V_0 \, \delta(\mathbf{r} - \mathbf{r}') \tag{5}$$

or the surface type,

$$V^{\delta}_{\mathrm{surf}}(\mathbf{r}, \mathbf{r}') = V_0 \, \left[ 1 - \frac{\rho(\mathbf{r})}{\rho_0} \right] \, \delta(\mathbf{r} - \mathbf{r}'), \tag{6}$$

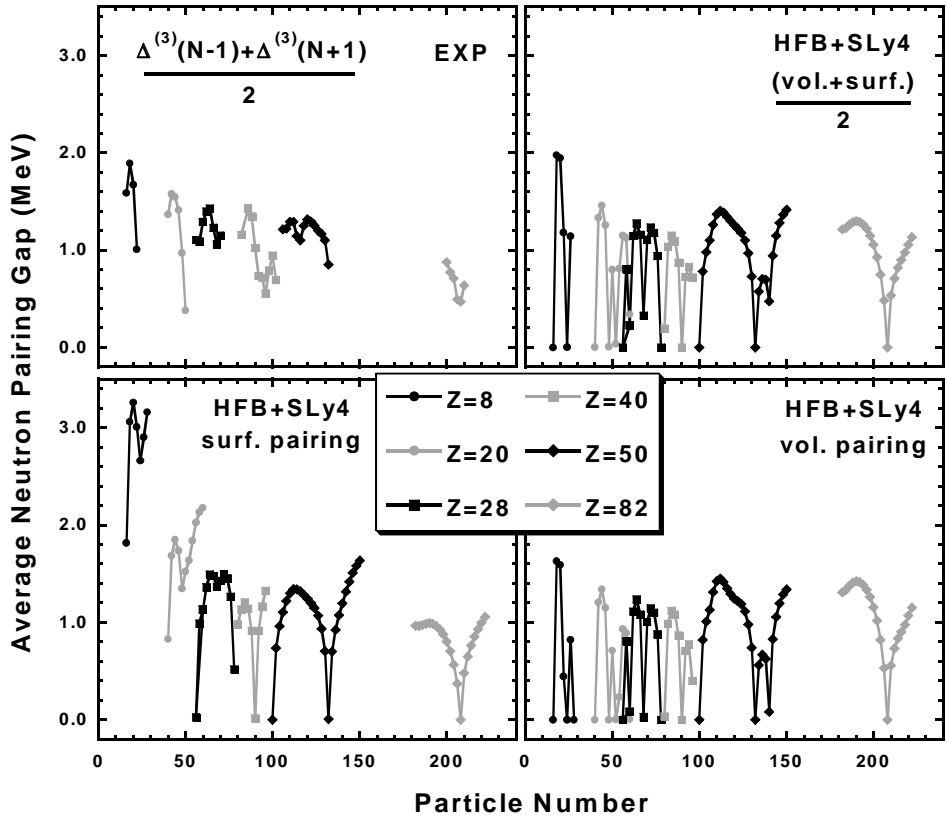

*Figure 1.* Comparison between the experimental staggering parameters (upper left panel) and the average neutron pairing gaps calculated within the spherical HFB method for the Skyrme SLy4 force and three different versions of the zero-range pairing interaction.

where $\rho_0$=0.16 fm$^{-3}$ is the saturation density, and $V_0$ defines the strength of the interaction. (The origin of the terms "volume" and "surface" has been discussed in Refs. [6, 28]. See Ref. [29] for more discussion on density dependence.) In our calculations, for every form of the pairing force, the value of $V_0$ is fixed by requiring that the experimental value of the neutron pairing gap in $^{120}$Sn (1.245 MeV) is reproduced within a given energy cut-off parameter $\bar{e}_{\max}$, cf. Ref. [30]. In fact, when using contact pairing forces, one should view the cut-off parameter to be an additional parameter defining the force. In the present study the value of $\bar{e}_{\max}$=60 MeV is used.

Under the conditions specified above, we have performed the spherical coordinate-space HFB calculations in semi-magic even-even nuclei. Results obtained for the average pairing gaps are shown in Figs. 1 and 2 for neu-trons and protons, respectively. In the upper left panels we show the values of experimental three-point staggering parameters $\Delta^{(3)}$ centered at odd particle numbers [31, 32] and averaged over the two particle numbers adjacent to the even value. The experimental data from the 1995 atomic mass evaluation [33] were used.

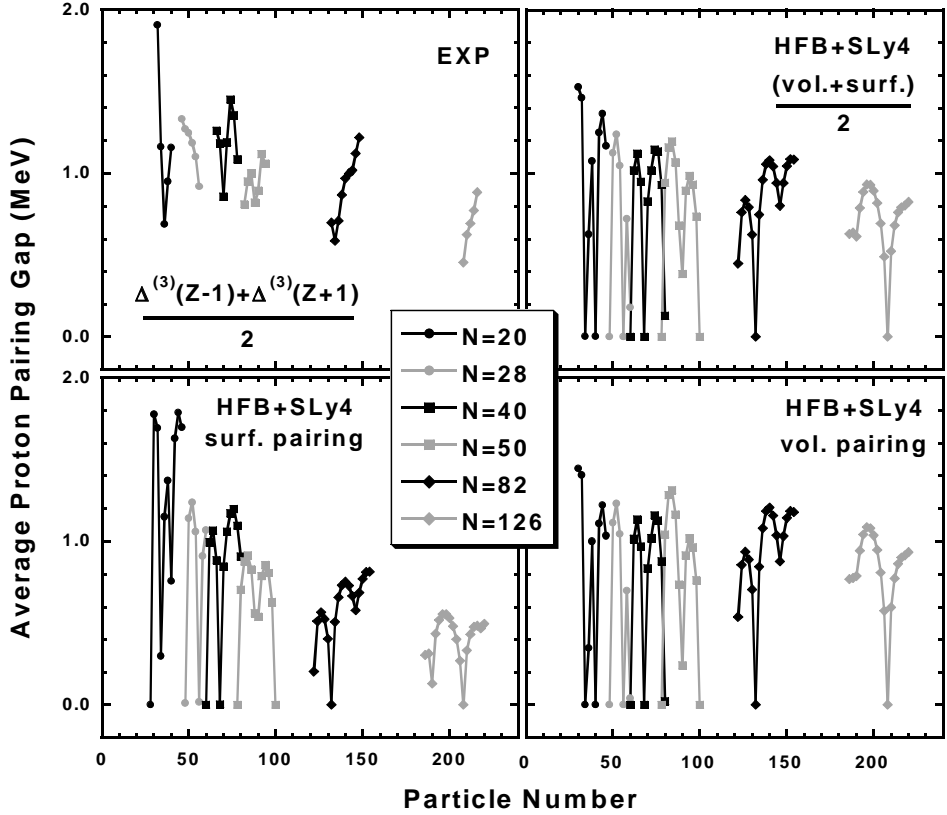

*Figure 2.* Same as in Fig. 1 except for the average proton pairing gaps.

The lower left and right panels in Figs. 1–2 show the results obtained for the surface and volume pairing interactions, respectively. When compared with the experimental numbers, one sees that both types of pairing interaction are in clear disagreement with experiment. The surface interaction gives the pairing gaps that increase very rapidly in light nuclei, while the volume force gives the values that are almost independent of $A$. The surface pairing in light nuclei is so strong that pairing correlations do not vanish in doubly magic nuclei such as $^{16}$O or $^{40}$Ca. The experimental data show the trend that is intermediate between surface and volume; hence, below we

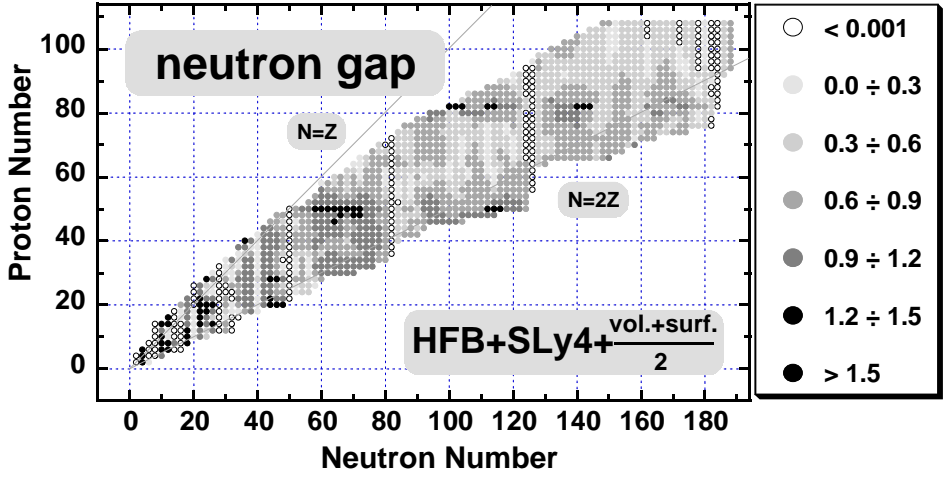

*Figure 3.* Average neutron pairing gap in even-even nuclei calculated within the deformed HFB+THO method for the Skyrme SLy4 force and contact mixed pairing interaction.

study the intermediate-character pairing force that is half way in between, i.e., it is defined as:

$$V_{\text{mix}}^{\delta}(\mathbf{r}, \mathbf{r}') = \frac{1}{2}\left(V_{\text{vol}}^{\delta} + V_{\text{surf}}^{\delta}\right) = V_0 \left[1 - \frac{\rho(\mathbf{r})}{2\rho_0}\right] \delta(\mathbf{r} - \mathbf{r}'). \qquad (7)$$

The upper right panels in Figs. 1 and 2 show the results obtained with the mixed pairing force. It can be seen that one obtains significantly improved agreement with the data, although a more precise determination of the balance between the surface and volume contributions still seems to be possible. One should note that similar itermediate-character pairing forces have recently been studied in Ref. [34] where it was concluded that pairing in heavy nuclei is of a mixed nature.

In Fig. 3 we present preliminary results for the average neutron pairing gaps calculated with the intermediate-character pairing force within the deformed HFB+THO method. Detailed analysis and discussion of these calculations will be presented in a forthcoming publication.

## 5. Acknowledgments

This work has been supported in part by the Bulgarian National Foundation for Scientific Research under project Φ-809, by the Polish Committee for Scientific Research (KBN), and by the U.S. Department of Energy under Contract Nos. DE-FG02-96ER40963 (University of Tennessee), and